\documentclass{nature}
\usepackage{graphicx}

\title{The One-Dimensional Wigner Crystal in Carbon Nanotubes}

\author{Vikram V. Deshpande$^{1}$ \& Marc Bockrath$^1$}

\begin{document}

\maketitle

\begin{affiliations}
\item Applied Physics, California Institute of Technology, Mail Stop 128-95, Pasadena, CA 91125
\end{affiliations}

\begin{abstract}
Electron-electron interactions strongly affect the behavior of low-dimensional systems. In one dimension (1D), arbitrarily weak interactions qualitatively alter the ground state producing a Luttinger liquid (LL)\cite{Lutreview} which has now been observed in a number of experimental systems\cite{changrev,598-601,273-276,gruner,yacobydoublewire}. Interactions are even more important at low carrier density, and in the limit when the long-ranged Coulomb potential is the dominant energy scale, the electron liquid is expected to become a periodically ordered solid known as the Wigner crystal\cite{PhysRev.46.1002}. In 1D, the Wigner crystal has been predicted to exhibit novel spin and magnetic properties not present in an ordinary LL\cite{PhysRevLett.71.1864,matveevwcprl,125416,195344,036809}. However, despite recent progress in coupled quantum wires\cite{113307,M.Yamamoto07142006}, unambiguous experimental demonstration of this state has not been possible due to the role of disorder. Here, we demonstrate using low-temperature single-electron transport spectroscopy that a hole gas in low-disorder carbon nanotubes with a band gap is a realization of the 1D Wigner crystal. Our observation can lead to unprecedented control over the behavior of the spatially separated system of carriers, and could be used to realize solid state quantum computing with long coherence times.
\end{abstract}

Carbon nanotubes are high mobility quantum wires that may enable the study of the intrinsic properties of the 1D electron gas without interference from disorder. Single quantum-dot transport experiments have been performed, demonstrating Coulomb blockade\cite{126801,sapmaz:153402} and Kondo physics\cite{342-346}, down to the few-electron-hole regime\cite{389,minotorbital,daicleantubes}. These experiments have generally been interpreted using the orthodox model and four-fold shell-filling. However, deviations have been noted at low density\cite{389,daicleantubes}, suggesting the importance of electron-electron interactions. In 1D, the ratio of inter-electron Coulomb energy to kinetic energy goes as $1/(na_B)$, where $n$ is the carrier density, $a_B=\epsilon \hbar^{2}/me^{2}$ is the Bohr radius, $\epsilon$ is the dielectric constant, $e$ is the electric charge, $\hbar$ is Planck's constant, and $m$ is the electron effective mass. On the basis of this order-of-magnitude estimate (see \textit{e.g.} ref. \citen{matveevwcprl}), the Coulomb energy should dominate and the system should begin to cross over to a strongly interacting regime at $n\sim {a_B}^{-1}= me^{2}/\epsilon\hbar^{2}$. While most previous studies in the few-electron regime have been done on nanotubes with small bandgaps ($<$100 meV), observing the Wigner crystal state is more favorable in nanotubes with larger gaps, where the effective mass is larger.

We report axial magnetic field measurements on low-disorder carbon nanotubes with larger bandgaps than previous studies. We observe for the first time three distinct regimes as a function of carrier density and magnetic field: (I) a completely spin and isospin polarized state, (II) an isospin polarized, spin antiferromagnetic state, and (III) an unpolarized state with a four-fold addition energy period. The transitions among these regimes can be quantitatively and intuitively explained using a Wigner crystal picture based on a gapped LL model\cite{levandtsv} with the carriers represented by spatially localized solitons. The value of the soliton width we obtain from our analysis, corresponding to the spatial extent of the carrier wavefunctions, agrees with theoretical estimates\cite{levandtsv,235428}. Surprisingly, we observe a sudden quenching of the Kondo effect for odd hole states as the magnetic field is tuned through the transition from the antiferromagnetically ordered regime to the four-fold filling regime. This can be understood based on the formation of large spin states with spin $S\geq 3/2$ resulting from the interplay between the spin and isospin degrees of freedom in the Wigner lattice.

Nanotubes with a bandgap are known to be more susceptible to disorder than metallic nanotubes\cite{5098-5101}. Our fabrication procedure, based on Cao \textit{et al.}\cite{daicleantubes}, is designed to reduce disorder by growing nanotubes in a suspended geometry to eliminate perturbations from the substrate and directly over metal contacts in order to avoid disorder due to chemical processing. Fig 1a shows a scanning electron microscope image of a representative device, while Fig 1b shows a schematic diagram of our devices, consisting of a nanotube with attached source, drain and gate electrodes. 

At low temperatures, we typically observe Coulomb peaks in the conductance $G$ versus the gate voltage $V_g$, when each additional hole is added to the nanotube dot. Fig 1c shows such $G$ versus $V_g$ data taken at temperature $T=1.4$ K from a device D1 with length 500 nm. $V_g>$1 V depletes the nanotube, placing the Fermi level in the bandgap. The regularity of the Coulomb peaks indicates that we are studying a single quantum dot that is essentially free from disorder and pinning. We also perform non-linear transport spectroscopy; Figure 2a shows a color-scale plot of $dI/dV$ versus source-drain bias $V$ and $V_g$ for D1. The data exhibits Coulomb blockade diamonds, shown by the dashed lines, and analysis of the data yields a charging energy $U\sim$10 meV, consistent with that expected for a 500 nm-long nanotube dot. We also measure the scale factor $\alpha$ between energy and gate voltage and find for D1 $\alpha\approx14$. Using the value for $\alpha$ and a measurement of $G$ versus $V_g$ over the bandgap, we infer the bandgap $2\Delta$, which for D1 was found to be $\sim$220 meV. 

Each added hole has both a physical spin and an isospin corresponding to the sense of the orbital motion of the carriers around the tube waist\cite{minotorbital}. In a magnetic field $B$, the ground state energy increase from each hole shifts by the sum of contributions $E_{orb}$ from its orbital magnetic moment\cite{minotorbital,daicleantubes} and the Zeeman energy $E_Z$ from the hole spin\cite{681-4}. These energies are given by $E_{orb}=\pm\mu_{orb}B=\pm rev_{F}B/2\approx0.41$meV$B$[T]$ r$[nm]\cite{minotorbital,daicleantubes}, and $E_{Z}=\pm g\mu_{B}B/2\approx\pm0.058B$[T] meV, where $\mu_{B}$ is the Bohr magneton, $r$ is the tube radius, $e$ the electric charge, $v_{F}$ the Fermi velocity, $g\approx2$ is the electron $g$-factor. As the position of $N$th Coulomb peak in $V_g$ is a factor $\alpha$ times the energy difference between the $N$ and $N+1$ hole ground state, we determine the spin and isospin for each added hole by applying $B$ parallel to the tube axis (Fig. 2b) and studying the Coulomb peaks' shifts versus $B$. Our results are plotted in Fig. 2c, which shows a color-scale plot of $G$ vs. $B$ and $V_g$ from D1.

Three distinct regimes are evident. The peaks up to number $\sim$12 are nearly parallel from near $B=0$ T up to $B=8$ T, shifting to larger $V_g$ for larger $B$ (positive slope). This region is referred to as region I. At more negative $V_g$, the slopes alternate between two different positive values (region II). Finally, a region where both positive and negative slopes are observed, as well as approximately horizontal features occurring quasi-periodically in $B$ (region III). The abrupt change of the slopes of the Coulomb peaks, particularly evident at the region II-III boundary indicates a change in the ground state wavefunction configuration. Curves are shown superposed on the graph approximately delineating the three regions corresponding to the results of a theoretical model fit that will be discussed below. 

Figure 3a shows the measured slopes at $B=8$ T in region I and II versus peak number. The peaks in region I have nearly the same slope, with $d(E_Z+E_{orb})/dB\sim 0.4$ meV/T, while peaks in region II show a pronounced alternation in addition energy, with an amplitude $\Delta E_s\sim 0.14$ meV/T, and generally falling below the data in region I. This indicates that in region I, all the holes enter the dot with the same spin and isospin. We thus find $\mu_{orb}\sim 0.33$ meV/T, yielding an estimated tube radius of $\sim$0.8 nm. In region II, the holes enter with alternating spins, but with the same isospin. From $\Delta E_s$ we estimate the electron $g$-factor to be $\approx$2.4, in good agreement with the theoretical value. Similar behavior has been found in all devices with a significant gap that showed regular Coulomb oscillations, attesting to the generality of this behavior in low-disorder samples.

The spin polarization observed in region I is surprising. In a shell-filling picture, the subbands corresponding to different isospin are split by the field\cite{minotorbital,daicleantubes}, and holes first begin to fill the lowest energy subband. This could account for the consistent isospin of the added holes. Yet, we expect because of the spin degenerate hole wavefunctions, they should enter with alternating spins except at sufficiently high magnetic field $B_c$ where the Zeeman energy would exceed the mean level spacing $\Delta E$. Taking a parabolic potential as in ref. \citen{389}, we find $\Delta E\sim1.5$ meV, giving $B_c=13$ T, precluding entirely the existence of region I over the range of magnetic fields studied. Using a hardwall potential, as done in ref. \citen{daicleantubes}, but with the additional inclusion of spin yields significantly fewer electrons in region I than we observe and also predicts for region II level crossings that would produce kinks in the evolution of the Coulomb peaks and singlet-triplet Kondo-valley conductance enhancements\cite{764-767}, which we do not observe. Thus, we cannot readily account for the observed behavior using simple shell-filling pictures, consistent with the strong interactions expected at these low densities. 
    
This behavior, however, can be naturally accounted for by assuming that the holes form a 1D Wigner crystal, in which the carriers become periodically ordered. Considering spin only, the inter-carrier exchange coupling $J$ \cite{matveevwcprl} is expected to be antiferromagnetic, as opposite spin carriers can both occupy their lowest energy molecular orbital. In nanotubes, the additional degree of freedom provided by the isospin results in modifications to the usual picture of exchange based on spin alone: the minimum energy state occurs when nearby carriers have different spin or isospin quantum numbers. Otherwise, the nearby carriers pay an energy cost $J$. When $n$ increases and the carrier wavefunctions overlap more strongly, a four fold addition energy period is expected; after exhausting all four spin and isospin combinations, adding the next carrier requires additional exchange energy.

When $B>0$ the state of the Wigner crystal is a competition between magnetic and exchange energy. Since $J$ is predicted to become exponentially small at low $n$\cite{195343,195344}, we expect at low $n$ the total energy will be minimized by having the holes both spin and isospin polarized. At intermediate $n$, since $E_{orb}>>E_Z$, we expect a transition to an antiferromagnetically spin ordered, isospin polarized chain when $g\mu_B B=2J$. For larger $n$, where non-nearest neighbor exchange becomes important, we expect to observe a four-fold addition energy pattern as discussed above. This picture thus qualitatively accounts for the existence and behavior of the three distinct regions as well as the transitions among them as $n$ increases. 

To quantitatively analyze our data, we utilize the theory of Levitov and Tsvelik\cite{levandtsv}. The carriers in the 1D Wigner crystal within the nanotube are treated in terms of a linear combination of a charged bosonic field and three neutral flavor bosonic fields, chosen to yield their given spin and isospin. The fields' behavior is governed by a kinetic energy cost for localization that competes with a gap-dependent potential energy gain. This yields spatially-localized soliton solutions for the fields, with an optimal width $w$ for the flavor solitons. Theoretical estimates yield a range of $w\sim5-20$ nm for sample D1\cite{235428,levandtsv}. Figure 3b shows the classical spatial variation of the four fields necessary to produce a chain with the four different spin and isospin combinations. 
    
We use the theory of Levitov and Tsvelik in a classical approximation (details are provided in the Supplementary Discussion) to compute the ground state energy for a spin polarized (I), antiferromagnetically ordered (II), and four-fold period (III) ground state using the spin and isospin configurations shown in the insets to Figs. 3a and 3b. As $n$ increases, the flavor solitons overlap more strongly, increasing $J$. Thus, we expect as $n$ increases the ground state for $B>0$ will make transitions among the three states. However due to $e.g.$ the non-uniform potential along the tube, we expect that the transitions between these states should happen via a sequence of individual spin or isospin flip events with a similar energy per carrier as calculated by our method. Using $w$ as a fitting parameter, we fit the observed transition between regions II and III to the calculated field where the two states are degenerate. From the data in Fig. 2c we find $w\sim9$ nm. This fit is superposed on the same data as the yellow curve and follows the observed boundary closely. The fitted value agrees well with the range of theoretically expected values. 

Using the fitted value for $w=9$ nm, the boundary between regimes I and II is computed directly, plotted in Fig. 2c as the red curve. This curve indicates the expected number of spin-polarized holes added at $B=8$ T is $\sim$9 in agreement with the observed number $\sim$12, although this boundary is not as distinct as that between II and III, which we attribute to thermal fluctuations that should blur the boundary by $\sim k_B T/g\mu_B\simeq 1$ T. Nevertheless, given that the curve is plotted without free parameters, the agreement is satisfactory.   

Finally, analysis of the transition between regions II and III reveals a unique aspect of Wigner crystallization. This is shown as a color-scale plot of the conductance, $G$ versus $B$ and $V_g$ in Fig. 4a. Regions II and III are shown, divided by a black dotted line. In region II, $G$ between adjacent Coulomb valleys alternates, with larger $G$ in the odd-hole valleys, as shown in Fig 4b. This is a well-known signature of the Kondo effect\cite{156-159}, which results from the screening of a local spin on the nanotube with the conduction electrons in the electrodes to form a spin singlet state. This state has characteristic binding energy $\sim k_BT_K$, where $T_K$ is the Kondo temperature\cite{156-159}. The conductance alternation between valleys, and slopes of the Coulomb peak motion discussed previously indicates that the spin of the nanotube changing between $S=S_0$ in the even valleys and $S=S_0+1/2$ in the odd valleys, where $S_0$ is a constant\cite{681-4}. 

Remarkably, when $B$ is tuned from region II to region III in an odd-hole valley (from A to B Fig. 4a), $G$ drops by $\Delta G\sim 1.5$ $\mu$S. In contrast, in the even-hole valley no such drop is observed (Fig. 4a points C and D). Fig. 4c shows $G$ versus $B$ taken along the red and yellow vertical dashed lines in adjacent Coulomb valleys. 

A shell-filling model could possibly account for the ridges by level crossings and an enhancement of $T_K$ because of the orbital degeneracy\cite{orbitalkondo}. However, this picture cannot readily account for the observed conductance drop $\Delta G$ since an odd hole state with total spin $S=S_0+1/2$ should remain in the same spin state even when the up and down isospin bands first align at the Fermi level, as illustrated in Fig. 4d. This picture therefore predicts $\Delta G=0$, contradicting the observed dip. 

In a Wigner crystal, the transition between region II and III occurs when $B$ is lowered so that a hole flips its isospin.  This hole also flips its physical spin to minimize its magnetic energy without an exchange energy cost. The total spin state becomes $S=S_0+3/2$, as shown schematically in Fig 4e. Since $T_K$ generally becomes lower as $S$ increases\cite{764-767}, $G$ should drop between A and B, as observed. Thus the observation of $\Delta G>0$ in the odd-hole valley provides a clear experimental signature of the 1D Wigner crystal state. Taken together with the quantitative agreement to the theory of Levitov and Tsvelik, our data provides convincing evidence that carriers in carbon nanotubes at low densities form a 1D Wigner crystal.

The realization of this long predicted state can now be used to test theories of interacting electrons in 1D in the clean limit\cite{matveevwcprl,125416,195344,036809,235428}. For instance, at $B = 0$ the equilibrium state is determined by a competition between the thermal energy $k_B T$ and $J$. If $k_B T >> J$, the spins and isospins can flip freely because of thermal fluctuations. This spin-incoherent regime is predicted to exhibit different behavior from an ordinary quantum wire, for example reduced conductance, which may be related to the 0.7 structure observed in quantum point contacts\cite{matveevwcprl}. 

Moreover, we note that at low density, the experiment achieves a carrier separation of $\sim$100 nm which gives experimental access to control 
of individual exchange couplings, say using local gates. One could then utilize the many-body Wigner crystal as a chain of quantum bits towards realizing spin-based quantum computing in carbon, where intrinsic spin lifetimes are expected to be longer than in conventional semiconducting materials.

\pagebreak
\noindent\textbf{\large Supplementary Discussion}

\noindent\textbf{Calculation of exchange energy}

\indent \indent The Levitov and Tsvelik theory\cite{levandtsv} treats the carriers in a nanotube as a gapped Luttinger liquid (LL). The interactions in the gapped LL are characterized by a charge stiffness parameter $K$ which is related to the inter-carrier interaction potential and can be usefully approximated by a constant $\sim20-40$\cite{levandtsv,235428}. The interactions renormalize the non-interacting electron bandgap, making it $\sim K^{1/2}$ times larger.

In a nanotube LL, the two spin and two isospin states are described by one charged and three neutral flavor bosonic fields. When a spectral gap is introduced into the LL, the minimum energy field conﬁguration becomes a lattice of sine-Gordon solitons, concentrating charge, spin, and isospin into localized packets to form the 1D Wigner crystal. The characteristic width $w_c$ of the charge solitons is related to the measured gap $\Delta$ by $w_c\sim\hbar v_c/\Delta$\cite{PhysRevB.52.10865}, where $v_c=K^{1/2} v_F$ is the charge mode velocity\cite{levandtsv}. The flavor soliton width is $w\sim K^{-1/2}w_c$\cite{levandtsv,235428}.

We compute the exchange energy in the Wigner crystal starting from the bosonized Lagrangian for the gapped Luttinger liquid with the four bosonic fields, which is the sum of two terms $V_0+V_1$. The competition between $V_0$ and $V_1$ leads to soliton solutions for these fields. The solutions\cite{levandtsv}, together with the potential, enable us to compute the energy treating the fields classically. 

The first term $V_0$ arises from the kinetic energy of localization\cite{Lutreview,levandtsv} 
\begin{equation}
V_0=\frac{\hbar v_F}{2\pi}\int dx \sum_{a=1}^3 (\partial_x\phi_a)^2
\end{equation}
where $\phi_0=\phi_{c+}$ is the charged field $\phi_{1,2,3}=\phi_{s+},\phi_{s-},\phi_{c-}$ are the three flavor fields, $q$ is a wavevector, $x$ is a position coordinate, $v_F$ is the Fermi velocity, and $\hbar$ is Plank's constant. The second term $V_1$ is an effective potential that arises from the gap, given by\cite{levandtsv} 
\begin{equation}
V_1=-4\lambda \int dx[\cos(\phi_{c+})\cos(\phi_{c-})\cos(\phi_{s+})\cos(\phi_{s+})+\sin(\phi_{c+})\sin(\phi_{c-})\sin(\phi_{s+})\sin(\phi_{s+})],
\end{equation}
Where $\lambda\simeq\hbar v_F/(4\pi w^2)$ is a (renormalized) coupling constant that is related to the physical flavor soliton width $w$. 

A single soliton centered at the origin for the charge field is approximated by\cite{levandtsv}

\begin{equation}
f_0(x)=\frac{\pi}{4} \left[e^{\frac{x}{\sqrt{K} w}} \theta (-x)+\left(2-e^{-\frac{x}{\sqrt{K} w}}\right) \theta (x)\right],
\end{equation}
where $\theta(x)$ is the unit step function. As a composite soliton of charge and flavor modes is the lowest energy state\cite{levandtsv}, the flavor fields then vary as
\begin{equation}
f_i(x)=\beta_i\frac{\pi}{4} \left[e^{\frac{x}{w}} \theta (-x)+\left(2-e^{-\frac{x}{w}}\right) \theta (x)\right],
\end{equation}
Where $\bar{\beta}$=(1,1,1) for a spin up, isospin up carrier, $\bar{\beta}$=(1,-1,-1) for a spin up, isospin down carrier, $\bar{\beta}$=(-1,-1,1) for a spin down, isospin up carrier, and $\bar{\beta}$=(-1,1,-1) for a spin down, isospin down carrier. A soliton lattice corresponding to a chain of $N$ carriers with density $n$ is given by
\begin{eqnarray}
\phi_0(x)&=&\sum_{j=1}^N f_0(x-j/n) \\
\phi_i(x)&=&\sum_{j=1}^N \beta_{ij} f_i(x-j/n),
\end{eqnarray}
where $\beta_{ij}$ are the numbers $\beta_i$ for the $j$th carrier. The exchange energy is computed by substituting the expressions for $\phi_0(x)$ and $\phi_i(x)$ into the the potential $V_0+V_1$ and computing the energy numerically. The magnetic energy is computed by multiplying the net spin by $g\mu_B B/2$ and the net isospin by $\mu_{orb} B$. The sum of the exchange and magnetic energy for a spin and isospin polarized electron or hole lattice, a spin-alternating, isospin polarized lattice, and a four-fold period lattice is used to determine which of these three states is the ground state. The result for the boundary between the four-fold and spin-alternating region is then converted into a mathematical interpolation function to fit to the data.  

We note that this model is expected to be valid as long as the soliton width significantly exceeds the tube diameter\cite{levandtsv}, as it is expected to theoretically, and as it does in our fitted results.


\textbf{$\!\!\!\!\!\!\!\!\!\!\!$References}
\vspace{2 mm}
\bibliography{reflib1}


\begin{addendum}
 \item We acknowledge Micro Nano Laboratory at Caltech and Nanotech at UCSB where fabrication was performed. We thank Chun Ning Lau, Gregory Fiete, Gil Refael, Ryan Barnett, Dmitry Novikov, Pablo Jarillo-Herrero, Charles Marcus, Bhaskaran Muralidharan, and H. Van der Zant for helpful discussions. We acknowledge the support of the Office of Naval Research and the Sloan foundation.  
 \item[Competing Interests] The authors declare that they have no
competing financial interests.
 \item[Correspondence] Correspondence and requests for materials
should be addressed to 

M.B.~(email: mwb@caltech.edu).
\end{addendum}
\begin{figure}
\caption{Experimental geometry and characteristic transport data. \textbf{a}, Scanning electron microscope image of a representative suspended nanotube device. Samples are fabricated by chemical vapor deposition growth of single-walled carbon nanotubes, from lithographically defined Fe-Mo catalyst islands\cite{kongsynth} across predefined Pt electrodes\cite{daicleantubes}, by flowing a gas mixture of methane (0.5 SLM) and hydrogen (0.7 SLM) at 800 $^\circ$C for 5 min. Devices are first studied at room temperature in an inert environment and the gate voltage and high bias (up to 1.5 V) electrical transport characteristics are recorded. Devices determined to have a bandgap and that show negative differential conductance at high bias with a maximum current of $\sim$10/$L$ $\mu$A (where $L$ is the known device length in $\mu$m), corresponding to individual, suspended, single-wall nanotubes\cite{155505}, are then selected for low-temperature measurement. \textbf{b}, Schematic of device geometry showing nanotube with attached source, drain, and doped Si gate electrode $\sim$600 nm below the nanotube. \textbf{c}, Conductance versus gate voltage for a 500 nm long device at temperature $T$=1.4 K.}
\end{figure}
\begin{figure}
\caption{Transport spectroscopy and magnetic field evolution of Coulomb peaks. \textbf{a}, Color plot of $dI/dV$ versus gate voltage $V_g$ and source-drain bias $V$. \textbf{b}, Magnetic field $B$ orientation relative to nanotube. \textbf{c}, Color plot of $G$ versus $V_g$ and magnetic field $B$. The three different regions and curves, obtained from a model fit, delineating them are described in the text. We expect similar behavior in the electron-doped regime due to electron-hole symmetry\cite{389}, but the Pt electrodes generally produce Schottky barriers for electrons that make it difficult to obtain very clear data for the electron-doped regime in larger gap devices. Nevertheless, such data should be possible to obtain in principle, perhaps using a different contact metal.}
\end{figure}
\begin{figure}
\caption{Energy shift of Coulomb peaks with $B$ and schematic diagram of solitons corresponding to the four combination of spin and isospin. \textbf{a}, Energy shift per tesla $dE/dB$ of Coulomb peaks measured at $B$=8 T in region I and II. $dE/dB$ is roughly constant in region I and shows alternation in region II. The value of the tube radius we find from analysis of this data is consistent with the range of radii typically reported for methane-based nanotube growth over the catalyst we employ\cite{kongsynth}, although knowledge of the tube radius is not essential to our analysis and interpretation. Note that the radius does not have to be related to the measured gap in general, since gaps can be opened due to radius-insensitive mechanisms like axial strains and twists\cite{minotorbital} and are expected to be enhanced due to interactions\cite{levandtsv,235428}. Note also that the results from region I differ from those of Tans \textit{et al.}\cite{761-764}, in which repeated addition of the same spin electron to a metallic nanotube was observed for $B<\sim 1$ T, followed by a complex spin-filling pattern at larger $B$. Here, the spin polarization is observed in a gapped nanotube only at low densities and over the entire measured range of $B$ up to 8 T.  Inset: schematic diagrams of spin configurations in region I and region II. The holes all have isospin up. \textbf{b}, Spatial variation of the four fields corresponding to charge sum in the two isospins ($\phi_{c+}$), and three flavor modes corresponding to the charge difference ($\phi_{c-}$), and spin sum ($\phi_{s+}$) and difference ($\phi_{s-}$). The corresponding hole configuration is shown at the top with the arrows indicating the spin and the color indicating the isospin, with blue indicating isospin up and red, hatched isospin down. The flavor soliton width $w$ is indicated by the red lines. The solitons are here depicted as non-overlapping. When there are $\sim L/w\sim 50$ holes in D1 the solitons should merge together. This corresponds to a nearly constant charge and flavor density. In this limit, we expect that the device should therefore cross over to a metallic hole-liquid like regime, where the behavior should be well-described by a conventional shell-filling picture, consistent with the well-defined four-fold periodicity of the Coulomb peaks in this regime.}
\end{figure}
\begin{figure}
\caption{Kondo effect and flavor configurations. \textbf{a}, Color scale plot of $G$ versus $V_g$ and $B$. The boundary between regions II and III is shown by the black dashed line. Points A and B are in an odd-hole valley, while points C and D are in an even-hole valley. \textbf{b}, $G$ versus $V_g$ showing Coulomb peaks in region II taken at $B=5$ T. \textbf{c}, Line traces from part \textbf{a} taken along the odd-hole valley (shown in red) and in the even-hole valley (shown in dark yellow). \textbf{d}, Shell filling picture of the transition between region II and III. Isospin up is shown by blue levels, while isospin down is shown by red levels. \textbf{e}, Wigner crystal picture of the transition between region II and III.}
\end{figure}

\pagebreak
\centering
\includegraphics{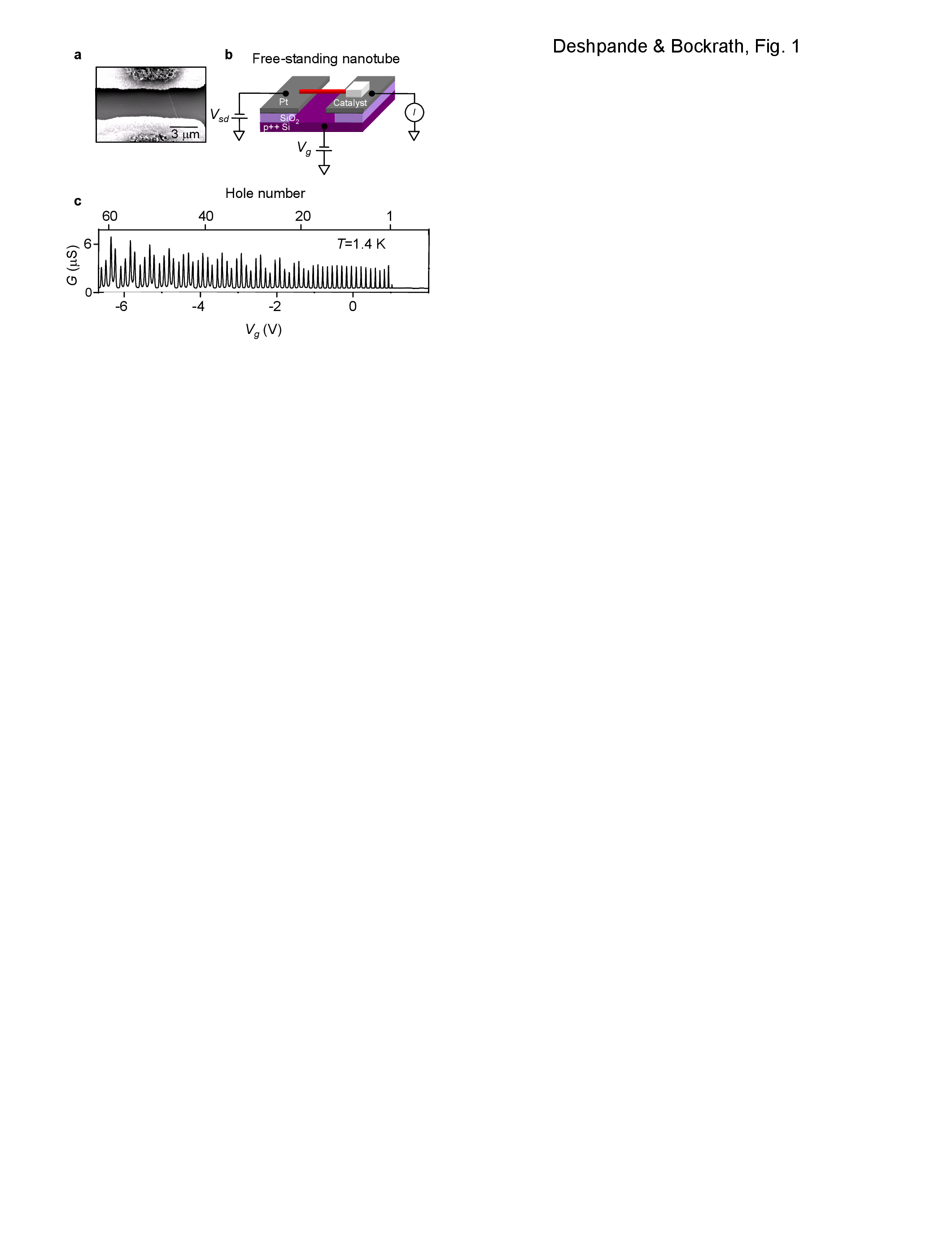}

\pagebreak
\centering
\includegraphics{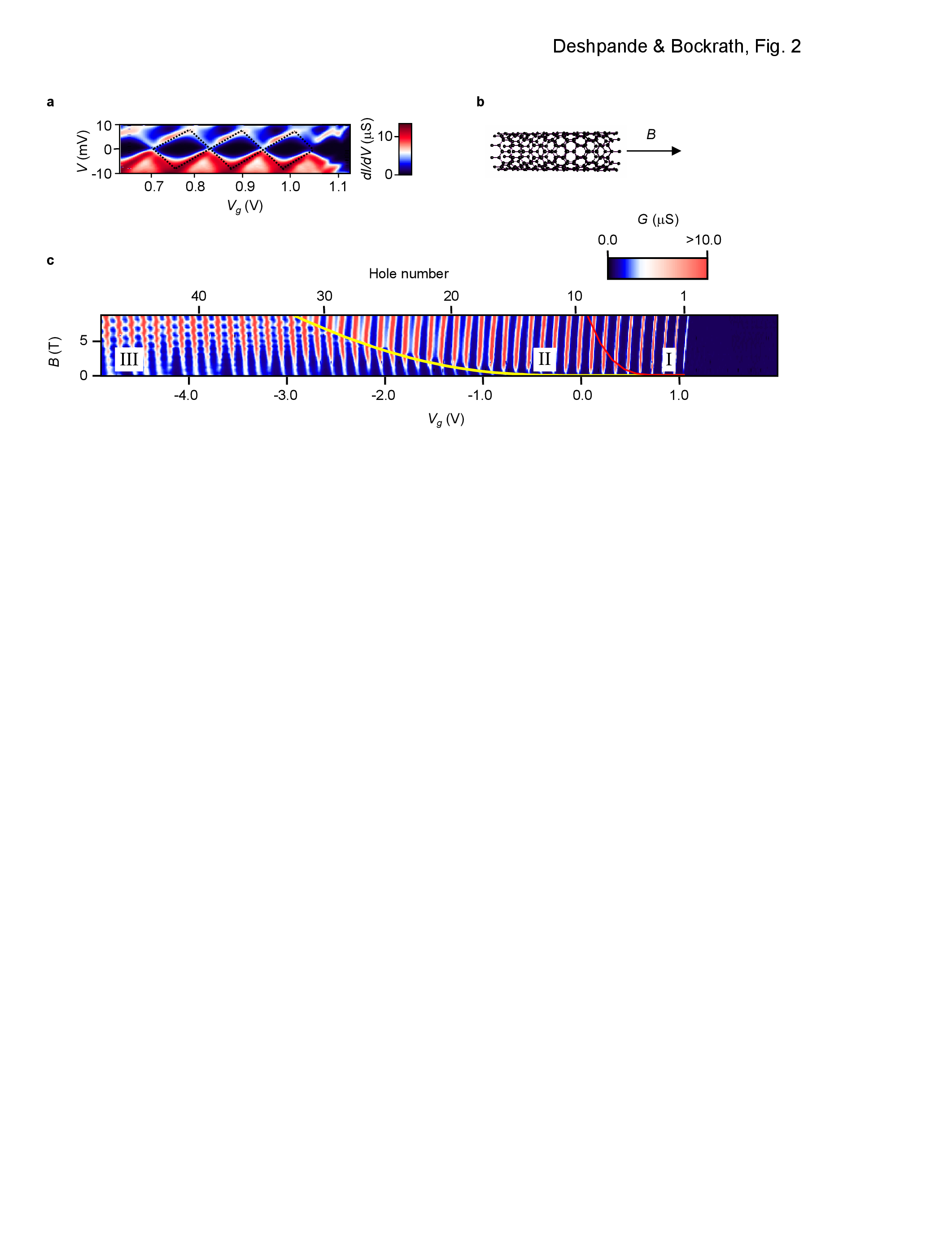}

\pagebreak
\centering
\includegraphics{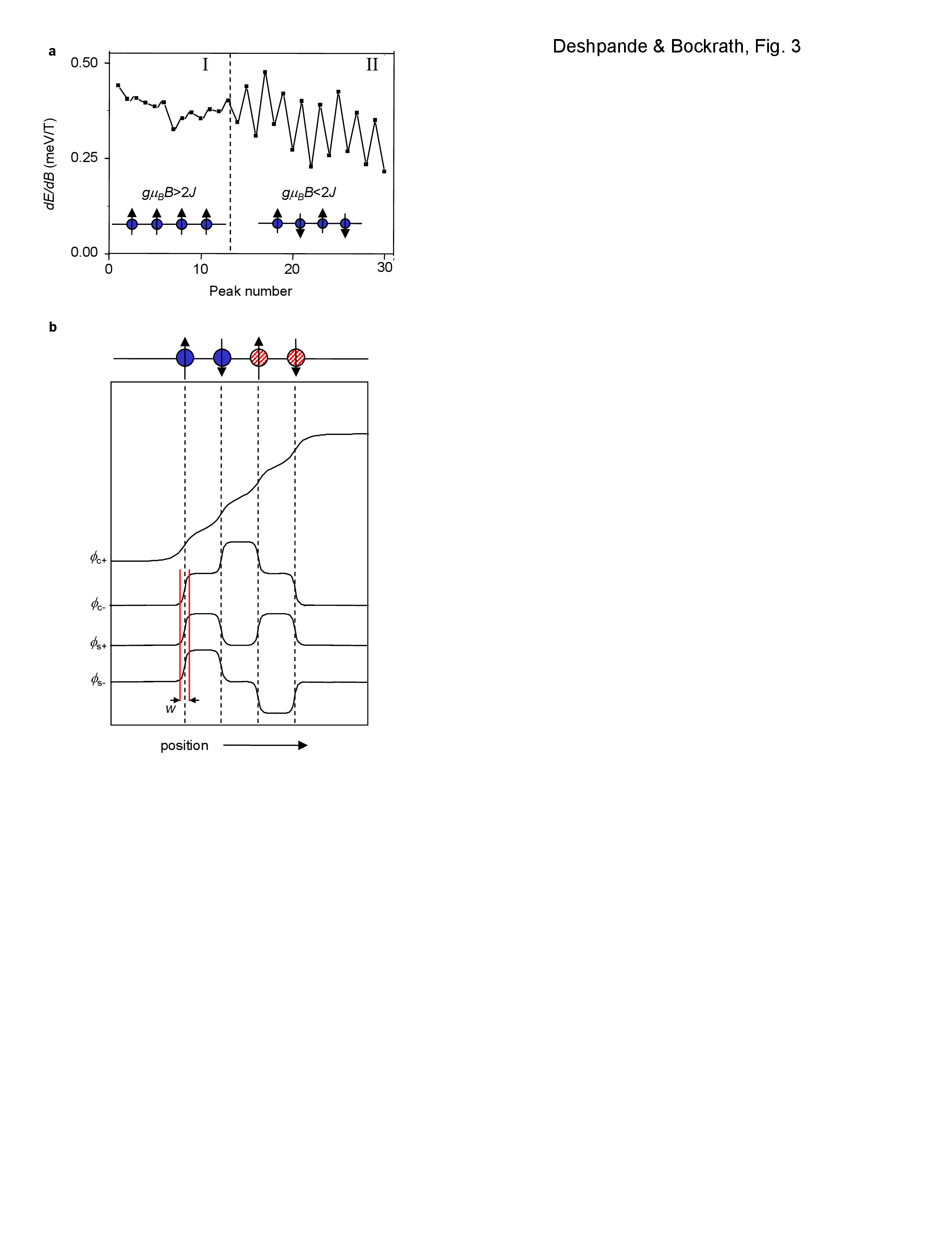}

\pagebreak
\centering
\includegraphics{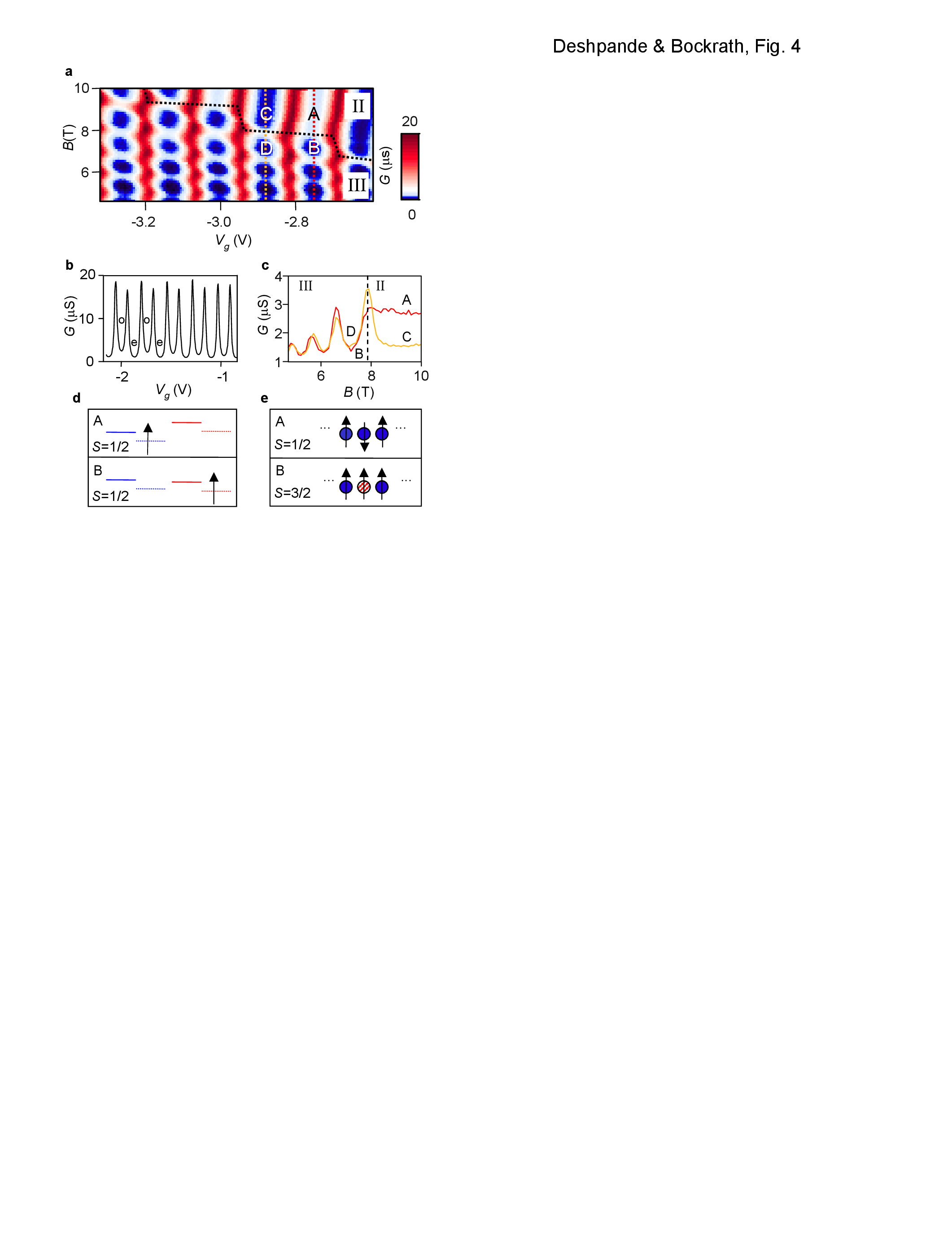}


\end{document}